\documentstyle[12pt,epsf]{article}
\title{
\vspace*{-1.3cm}
\begin{flushright}
{\normalsize Fermilab Conf-99/222-T}
\end{flushright}
\vspace{0.8cm}
Lattice determinations of the strange quark mass\thanks{Talk presented at KAON'99, University of Chicago, June 1999}}
\author{Sin\'ead Ryan\\ \begin{small}{\it Fermi National Accelerator Laboratory, Batavia Illinois, U.S.A.}\end{small} }
\date{ }
\begin{document}
\maketitle
\section{Introduction}
The importance of the strange quark mass, as a fundamental parameter of the Standard Model (SM) and as an input to many interesting quantities, has been highlighted in many reviews, eg in Ref~\cite{guido_lat98}. 
A first principles calculation of $m_s$ is possible in lattice QCD but 
to date there has been a rather large spread in values from lattice calculations.
This review aims to clarify the situation by explaining the particular systematic errors and their effects and illustrating the emerging consenus.

In addition, a discussion of the strange quark mass is timely 
given the recent results from KTeV~\cite{ktev} and NA48~\cite{na48} for $\epsilon^\prime /\epsilon$ which firmly establish direct CP-violation in the SM and when combined with previous measurements give a world average $\epsilon^\prime /\epsilon = (21.2\pm 2.8)\times 10^{-4}$.
This is in stark disagreement with the theoretical predictions which favour a low $\epsilon^\prime /\epsilon$~\cite{buras_review_and_others}. 

Although in principle $\epsilon^\prime /\epsilon$ does not depend directly on $m_s$ in practice it has been an input in current phenomenological analyses. This dependence arises because the matrix elements of the gluonic, $\langle Q_6\rangle_0$, and electroweak, $\langle Q_8\rangle_2$, penguin operators\footnote{keeping only the numerically dominant contributions for simplicity} 
are of the form $\langle\pi\pi |Q_i|K\rangle$ and final state interactions make them notoriously difficult to calculate directly. 
They have been, therefore, parameterised in terms of bag parameters, ${\cal B}_i$, the strange quark mass, $m_s$ and the top quark mass, $m_t$, as discussed in detail in Ref.~\cite{buras_review_and_others}. A recent review of lattice calculations of the matrix elements is in Ref.~\cite{guido}.
In this talk I will focus on some recent and careful lattice determinations of $m_s$, illustrating the reasons for the large spread in earlier results.
\section{The stange quark mass from lattice QCD}
In lattice QCD, $m_s$ is determined in two ways, each of which relies on 
calculations of experimentally measured quantities to fix the lattice bare coupling and quark masses. 
The 1P-1S Charmonium splitting, $M_\rho$ and $r_0$ are some of the parameters typically chosen to fix the inverse lattice spacing, $a^{-1}$.
To determine $m_s$ either $M_K$ or $M_\phi$ is used. It is an artefact of the quenched approximation that $m_s$ depends on the choice of input parameters, so that some of the spread in answers from lattice QCD can be attributed to different choices here. 
Naturally, some quantities are better choices than others being less sensitive to quenching or having smaller systematic errors.

The quark mass can be determined from hadron spectroscopy, using chiral 
perturbation theory to 
match a lattice calculation of $M_K$ (or $M_\phi$) to its experimental value with 
\begin{equation}
M_{PS}^2 = B_{PS}\frac{(m_i +m_j)}{2} + \ldots \;\; \mbox{{\bf or}}\;\; M_V = A_V + B_V\frac{(m_i+m_j)}{2} + \ldots 
\end{equation}
This is the hadron spectrum or vector Ward identity (VWI) method.

Alternatively, the axial Ward identity (AWI): 
$\partial_\mu A_\mu (x) = (m_i +m_j)P(x)$ imposed at quark masses to correspond to either the experimentally measured $M_K$ or $M_\phi$ determines $m_s$. 

The lattice bare masses and matrix elements are related to their continuum counterparts, in say the $\overline{MS}$ scheme, by the renormalisation coefficients, $Z_s$ or $Z_{(A,P)}$, calculated perturbatively or nonperturbatively,
\begin{eqnarray*}
m_s^{\overline{MS}}(\mu ) = Z_s^{-1}(\mu ,ma)m^0_q \; &,&\; (m_s+\overline{m})^{\overline{MS}}(\mu ) = \frac{Z_A(ma)}{Z_P(\mu ,ma)}\frac{\langle\partial_\mu A_\mu J(0)\rangle}{\langle P(x)J(0)\rangle}.
\end{eqnarray*}
$m_s$ has been calculated in all three lattice fermion formalisms: Wilson, staggered and domain wall. Although the domain wall fermion results are extremely interesting, since this approach has the good flavour structure of Wilson fermions while preserving chiral symmetry, the results for $m_s$ are still preliminary so I will focus on results with Wilson and staggered fermions. A description of the domain wall formalism and results can be found in Ref.~\cite{wingate}.

Comparing results from these different methods provides a nice check of lattice calculations. 

\section{Main uncertainties in the calculation}
The difference in early lattice results can be understood in terms of the treatment of systematic uncertainties in these particular calculations. The largest
of these are discretisation errors, calculation of renormalisation coefficients
and the quenched approximation.

\underline{{\it 1. Discretisation Errors :}}
The Wilson action has discretisation errors of ${\cal O}(a)$, so for a reliable result one needs fine lattices and a continuum extrapolation, $a\rightarrow 0$. See Figure~\ref{cp-pacs_np_ms} for the CP-PACS collaboration's quenched Wilson results~\cite{cppacs-wilson}.
The Sheikholeslami-Wohlert (SW) clover action includes a term $\sim c_{SW}\overline{\Psi}\sigma_{\mu\nu}F_{\mu\nu}\Psi$ and discretisation errors start at ${\cal O}(\alpha_sa)$, when $c_{SW}$ is determined perturbatively. 
The remaining $a$-dependence must be removed by continuum extrapolation, but 
the slope of the extrapolation is milder~\cite{fnal_ms}. 
A nonperturbative determination of $c_{SW}$~\cite{Alpha_npCsw} gives an ${\cal O}(a)$-improved action, which should futher reduce the lattice 
spacing dependence. Recent results from the APE, ALPHA/UKQCD and QCDSF collaborations use this approach~\cite{APE_NP,ALPHA_NP,QCDSF_NP}. The latter two groups include continuum extrapolations and find significant $a$-dependence ($\approx 15\%$ between the finest lattice and $a=0$ as found by ALPHA/UKQCD). 
In the case of the more commonly used VWI approach the slope of the extrapolation in $a$ is positive and therefore, $m_s$ at finite lattice spacing is too high, even with improvement.

The staggered fermion action is ${\cal O}(a)$-improved so the lattice spacing dependence should be mild. 

\underline{{\it 2. Renormalisation coefficients :}}
$Z_S$ and $Z_{(A,P)}$ can be determined perturbatively or nonperturbatively. A nonperturbative calculation is preferable as it removes any perturbative ambiguity. This was pioneered by the APE and ALPHA groups~\cite{Alpha_npCsw,Ape_npCsw}. 

For Wilson fermions perturbative corrections are smaller and therefore more reliable in the VWI approach (ie.~for $Z_S$) than in the AWI approach. In Ref.~\cite{APE_NP} the difference between nonperturbative results and boosted perturbation theory is $\sim 10\%$ for $Z_S$ and $\sim 30\%$ for $Z_P$ at $a^{-1}\sim 2.6$ GeV.
For staggered fermions the perturbative coefficients are large and positive
so the results are unreliable and nonperturbative renormalisation is 
essential. 
The perturbative staggered results are therefore too low and this effect combined with the too high values of $m_s$ from Wilson results at finite lattice spacing explain much of the spread in lattice results.

\underline{{\it 3. Quenching :}} 
Most calculations are done in the quenched approximation - neglecting internal quark loops - as a computational expedient. An estimate of this approximation, based on phenomenological arguments, was made in~\cite{fnal_ms}. The authors estimated that unquenching lowers $m_s$ by $\approx 20-40\%$. They also argued that $M_K$ rather than $M_\phi$ is a better choice of input parameter since it is less sensitive to quenching. Unquenched calculations by CP-PACS have shown that these estimates were of the correct size and sign~\cite{CPPACS_nf2}.

A number of clear trends are therefore identified:
\begin{itemize}
\item There is significant $a$-dependence in the Wilson action results which raises $m_s$ at finite lattice spacing. Although this is milder for the improved actions it is still present, as pointed out in Refs.~\cite{fnal_ms,lanl_ms,ALPHA_NP}. 
\item Using perturbative improvement, the VWI and AWI methods differ at finite lattice spacing but agree after continuum extrapolation. This indicates the methods have discretisation errors larger than the perturbative uncertainty. Nonperturbative renormalistion has a larger effect on AWI results, bringing them into agreement with VWI results at finite lattice spacing. However, discretisation errors remain a significant uncertainty and without a continuum extrapolation lead to an overestimate of $m_s$.
\item Perturbative renormlisation of staggered fermions results in an 
underestimate of $m_s$. Nonperturbative renormalisation is essential.
\item A lower value of $m_s$ is expected from an unquenched calculation.
\end{itemize}
\section{Recent results for $m_s$}
The systematic uncertainties in the lattice determination of $m_s$ are now well understood. Some recent results which I believe provide a definitive value of $m_s$ in quenched QCD and an unquenched result are now discussed. 
\subsection{Quenched results}
Table~\ref{ms_np_tab} compares a number of recent calculations of 
$m_s$. The JLQCD, ALPHA/UKQCD and QCDSF groups have removed all uncertainties within the quenched approximation.
JLQCD use staggered fermions and nonperturbative renormalisation~\cite{JLQCD_NP}. They observe mild $a$-dependence, as 
expected and take the continuum limit. The effect of nonperturbative renormalisation is considerable, again as expected: $\sim +18\%$ when compared to the perturbative result. 

The ALPHA/UKQCD collaborations~\cite{ALPHA_NP} and QCDSF~\cite{QCDSF_NP} use a nonperturbatively improved SW action and renormalisation and include a continuum extrapolation.
This explains the difference between their results and that of the APE group
(which has not been extrapolated to $a=0$). 
Interestingly, ALPHA/UKQCD, QCDSF and the Fermilab~\cite{fnal_ms} and 
LANL~\cite{lanl_ms} results for $m_s$ are in agreement.
The difference in analyses is nonperturbative versus perturbative renormalisation, indicating that the perturbative result for the VWI method is reliable (for Wilson fermions).
\begin{table}[h]
\begin{center}
\vspace{-0.5cm}
\begin{tabular}{|c|ccc|}
\hline
GROUP & \# lattice spacings & $a\rightarrow 0$ & $m_s^{\overline{MS}}$(2GeV) \\
\hline
APE '98~\cite{APE_NP}   &  2 & NO & 121(13) \\
\hline
FNAL '96~\cite{fnal_ms}	&   3 & YES & 95(16)\\
LANL '96~\cite{lanl_ms} & 3 & YES & 100(21)(10)\\
\hline
JLQCD '99~\cite{JLQCD_NP} & 4 & YES & 106(7.1) \\ 
QCDSF '99~\cite{QCDSF_NP} &  3 & YES                     & 105(4)\\
ALPHA/UKQCD '99~\cite{ALPHA_NP} & 4 & YES               & 97(4)\\ 
\hline
\end{tabular}
\label{ms_np_tab}
\caption{Quenched lattice results, the APE result is obtained at $a^{-1}$=2.6GeV}
\end{center}
\vspace{-1cm}
\end{table}
\subsection{An unquenched result}
\begin{figure}[ht]     
\begin{center}
\setlength{\unitlength}{1cm}
\setlength{\fboxsep}{0cm}
\begin{picture}(13,9.5)
\put(1,2){\begin{picture}(7,7)\put(-0.9,-0.4)
{\epsfxsize=6.5cm\epsfbox[10 30 560 590]{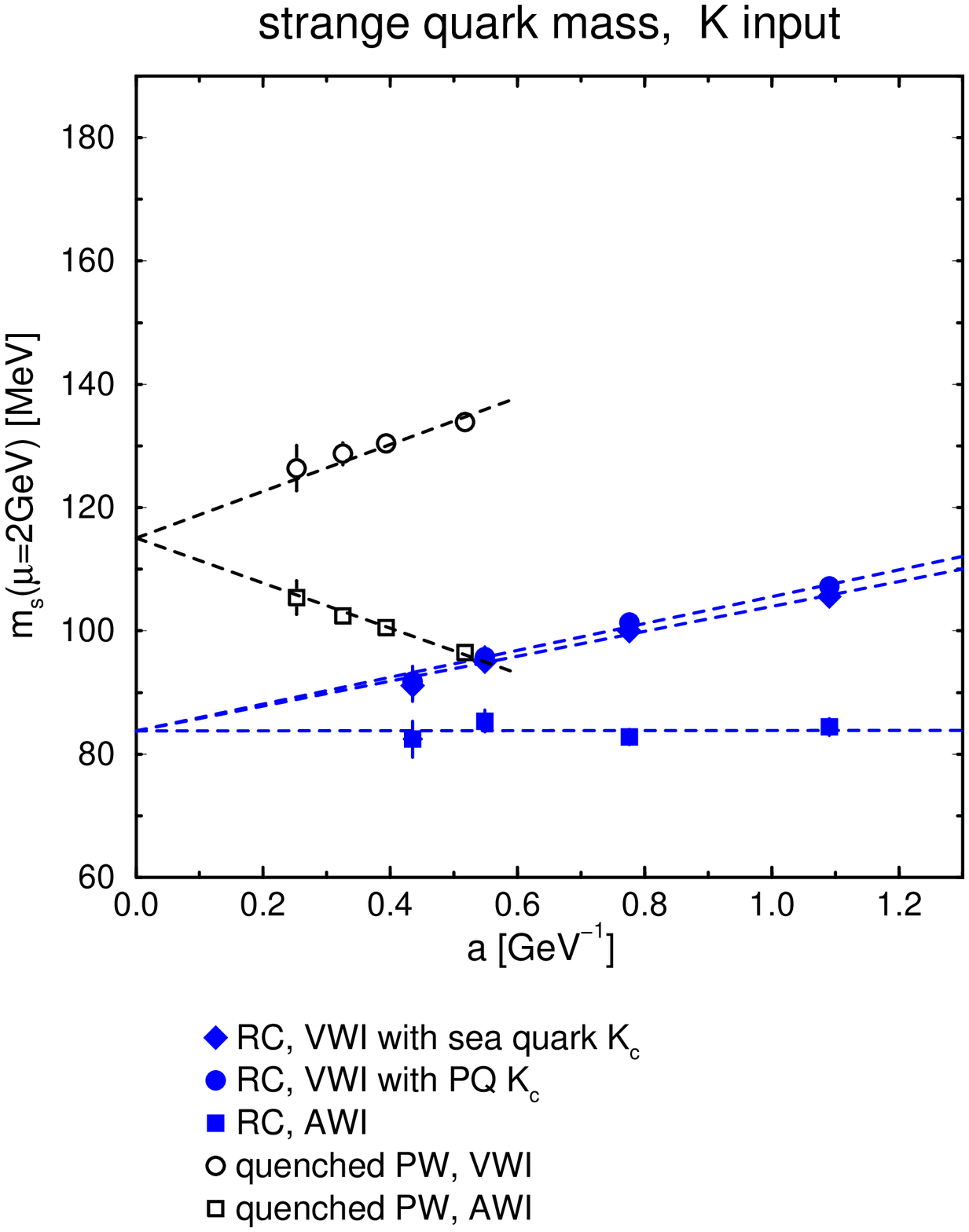}}\end{picture}}
\put(7.5,2){\begin{picture}(7,7)\put(-0.9,-0.4)
{\epsfxsize=6.5cm\epsfbox[10 30 560 590]{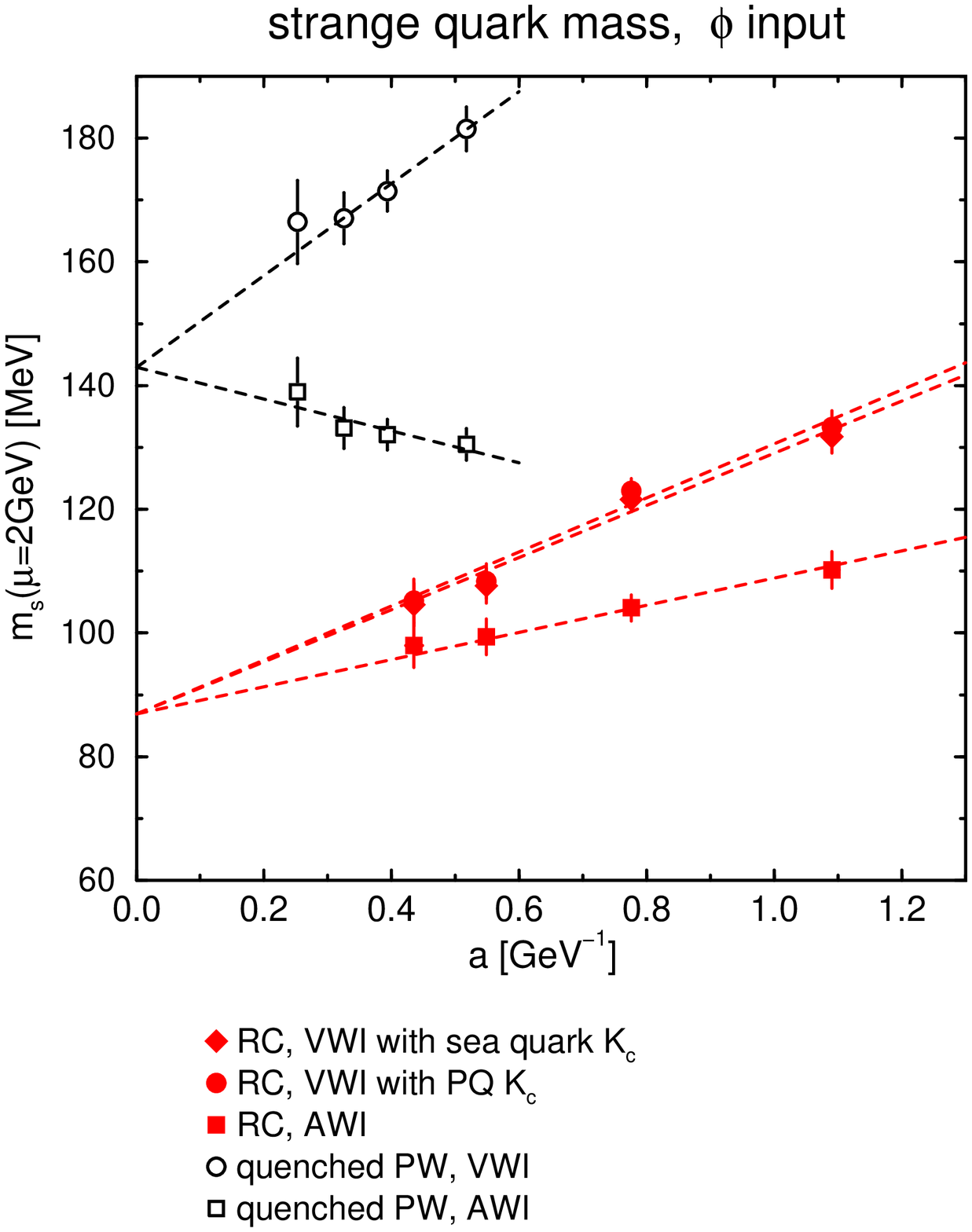}}\end{picture}}
\end{picture}
\end{center}
\vskip -3.0 cm
\caption[]{
\label{cp-pacs_np_ms}
\small RC is the RG-improved gluon and Clover action (with two definitions of quark mass in the chiral limit), PW are the quenched Wilson results already discussed. }
\vspace{-0.2cm}
\end{figure}

There are a number of new preliminary unquenched calculations of $m_s$~\cite{MILC} however, CP-PACS have recently completed their analysis~\cite{CPPACS_nf2}, shown in Figure~\ref{cp-pacs_np_ms}, so I will concentrate on this.
Since unquenching requires a huge increase in computing time it is prudent to use coarser (less time consuming) lattice spacings. This in turn requires improved actions to control the discretisation effects. CP-PACS use a perturbatively improved quark and gluon action
and extrapolate to the continuum limit.
The perturbative renormalisation is reliable with a remaining perturbative error of ${\cal O}(\alpha_s^2 )$, for Wilson fermions. The final result is $m_s^{\overline{MS}}(2\mbox{GeV}) = 84(7) \mbox{MeV}$.
Although this result disagrees with bounds derived from the positivity of the spectral function~\cite{laurent} it remains unclear at what scale, $\mu$, perturbative QCD and thus the bound itself becomes reliable.
CP-PACS conclude that unquenching lowers $m_s$ - compare the filled and open symbols in Figure~\ref{cp-pacs_np_ms}.
As in the quenched case the VWI and AWI methods differ at finite lattice spacing but extrapolate to the same result - compare the \begin{small}$\bigcirc$\end{small} and $\Box$ symbols. Finally, the strange quark mass obtained from the K and $\phi$ mesons yields consistent continuum values in full QCD: 84(7) MeV and 87(11) MeV respectively.
\section{Conclusions}
There has been much progress this year in lattice calculations of $m_s$. 
Current computing power and theoretical understanding are sufficient to determine $m_s$ to great precision.
A calculation removing {\it all} uncertainties would include unquenched simulations, a continuum extrapolation and nonperturbative renormalisation and can be done in the short term.
Simulations at $n_f=2$ and $4$ with an interpolation to $n_f=3$ are also a possibility.

\begin{figure}[h]     
\centerline{\epsfxsize=3.6in\epsfysize=2.6in\epsfbox{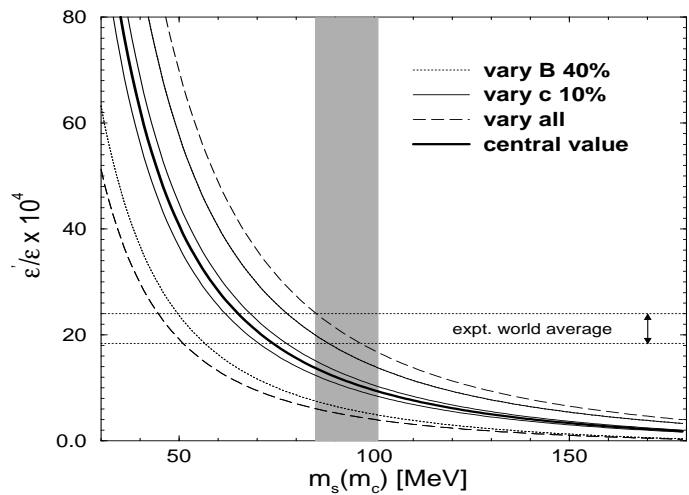}}
\caption{$\epsilon^\prime /\epsilon$ as a function of $m_s$ from Equation.~\ref{eprime_eqn}.}
\label{eprime_on_e}
\end{figure}

Finally, I look at the implications for $\epsilon^\prime /\epsilon$ from current theoretical calculations given the recent lattice calculations of $m_s$. 
The dependence is shown in Figure~\ref{eprime_on_e} from the analytic expression
\begin{equation}
\epsilon^\prime /\epsilon = \mbox{Im}\lambda_t\cdot \left [ c_0+\left ( c_6{\cal B}_6^{(1/2)} + c_8{\cal B}_8^{(3/2)}\right )\left (\frac{M_K}{m_s(m_c )+m_d(m_c )}\right )^2\right ] \label{eprime_eqn}
\end{equation}
and input from lattice calculations for ${\cal B}_i$~\cite{rajan_rev}. The values of other SM parameters are from Ref.~\cite{buras_jamin_laut}.
The lines represent the effect of varying the bag parameters and/or the Wilson coefficients and the band is the unquenched $m_s$ from CP-PACS, run to $m_c$. 
Further reducing the uncertainty on $m_s$ is more straightforward than for the ${\cal B}_i$ and can constrain theoretical calculations of $\epsilon^\prime /\epsilon$. Clearly the lower values of $m_s$ give higher $\epsilon^\prime /\epsilon$ values, in better agreement with experiment! 
%

%
\end{document}